# FREQUENCY JUMP AT LOW ENERGIES

Chuan Zhang[#], GSI Helmholtz Center for Heavy Ion Research, Planckstr. 1, Darmstadt, Germany
Holger Podlech, Institute for Applied Physics, Goethe-University, Frankfurt a.M., Germany

*Abstract*

One or more radio-frequency jumps are usually necessary for realizing a ≥100 AMeV/u proton or ion driver linac. Typically, such jumps happen in the range of $\beta$ = 0.2-0.6 between the resonator structures fitting to this $\beta$-range, e.g. DTL, HWR, CCL or elliptical cavities. We propose to perform the first frequency jump already at low energies ($\beta \leq 0.1$) between two RFQ accelerators, which can bring some unique advantages. First studies have been performed and the results proved that this idea is feasible and promising. Many efforts have been and are being made to address the most critical issue for the jumps i.e. the beam matching at the transition.

## INTRODUCTION

There is worldwide a common interest to develop intermediate- and high-energy (≥100 AMeV/u) proton or ion driver linacs for various modern applications. For such large-scale linacs, e.g. SNS [1], J-PARC [2], ESS [3], FRIB [4], MYRRHA [5], one or more radio-frequency jumps (RFJ) are usually required in order to keep the whole facility compact. Typically, the first RFJ happens in the range of $\beta$ = 0.2-0.6 between the resonator structures fitting to this $\beta$-range, e.g. DTL, HWR, CCL or elliptical cavities.

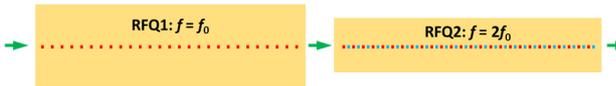

Figure 1: Schematic plot for the radio-frequency jump between two RFQ accelerators.

We propose to perform the first RFJ using the RFQ accelerator which is the standard accelerating structure for the low-energy end ($\beta$ = 0.01 to 0.1). Figure 1 is a schematic plot for the proposal. A similar idea to perform an RFJ between two RFQs was introduced in [6]. The differences between the two proposals are as follows:

- The motivation of that idea was to increase the current limit by taking advantage of the low frequency (more advantages at low frequencies can be found in [7]), but our goal is to reach the compactness.
- In that idea, a debunching section is needed as a transition of the two RFQs, but in our case, the beam will be kept bunched between two RFQs.

[#] c.zhang@gsi.de

The advantages that will be provided by our proposal can be roughly summarized as follows:
- The whole facility can reach a relatively higher frequency already at an early stage, so it can become even more compact.
- The RFJ will cause also a corresponding jump in the phase spread of the accelerated beam naturally. The RFQ accelerator can offer a much larger phase acceptance than other accelerating structures, and more important, it can provide a progressive bunching that will be very helpful to minimize the beam quality degradation due to the RFJ.
- The first drift tube of a DTL has to have a reasonable length to be built, but the RFQ accelerator has not such a limitation. Therefore, the RFQ-RFQ combination is the most promising RFJ solution at low energies.

But there are also some challenges to apply this method:
- The energy half-width of the separatrix is given by the following formula [8]:

$$w_{\max} = \sqrt{\frac{2qE_0 T \beta_s^3 \gamma_s^3 \lambda}{\pi mc^2}} (\varphi_s \cos\varphi_s - \sin\varphi_s) \qquad (1)$$

where $q$ is the charge, $E_0$ is the average electric field, $T$ is the transit time factor, $\beta_s$ is the normalized velocity, $\gamma_s$ is the Lorentz factor, $\lambda$ is the wavelength, and $\varphi_s$ is the synchronous phase, respectively. It can be concluded that an RFJ at low energies could lead to very likely a small energy acceptance.

- Typically, at the end of an RFQ accelerator, the accelerated beam is focused in one transverse plane but defocused in another one. However, the RFQ accelerator prefer to have an input beam that is focused in the both transverse planes. Therefore, how to match the beam transversely between two RFQs will be a difficult issue.

In this paper, the design concepts adopted for solving both longitudinal and transverse beam matching problems caused by the RFJ as well as the corresponding beam dynamics simulation results will be presented.

## DESIGN & SIMULATION

The RFJ design studies have been performed based on an RFQ-RFQ combination whose basic parameters are described in Table 1, where the inter-vane voltage values of the two RFQs are relatively moderate to keep their Kilpatrick factors lower than 1.5 so that they can be also feasible for CW operation. In addition, a 5 emA proton beam has been assumed as the input beam.

Table 1: Basic Parameters of the RFQ-RFQ Combination

|  | RFQ1 | RFQ2 |
| --- | --- | --- |
| $f$ [MHz] | 176 | 352 |
| $W_{in}$ [MeV] | 0.03 | 1.00 |
| $W_{out}$ [MeV] | 1.00 | 2.50 |
| $U$ [kV] | 46 | 55 |
| $L$ [m] | ~3 | ~3 |

In Fig. 2, the evolution of the main parameters of the two RFQs can be seen, where $a$ is the minimum electrode aperture, $m$ is the electrode modulation, and $\varphi_s$ is the synchronous phase, respectively.

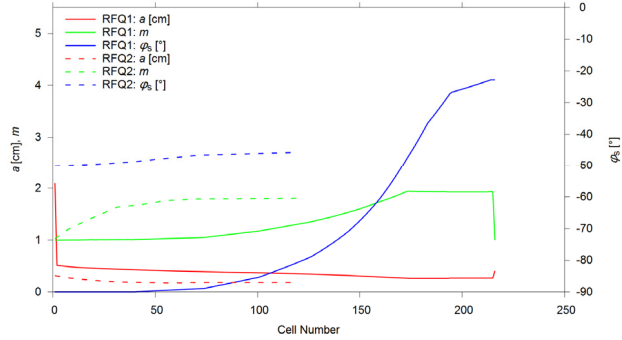

Figure 2: Evolution of main parameters along the RFQs.

In order to minimize the output energy spread at the RFQ1 exit for a better longitudinal matching into the RFQ2, the so-called New Four-Section Procedure (NFSP) [9, 10] that can provide a balanced and efficient beam bunching at low energies was adopted for the RFQ1 design. Figure 3 shows the simulated RMS (Root Mean Square) phase- and energy-spread along the accelerating channel. The particle dynamics simulation has been performed using the DYNAC code [11, 12] using a 4D-Waterbag input distribution with an emittance value of $\varepsilon_{in,n.,rms}$ = 0.2 $\pi$ mm-mrad.

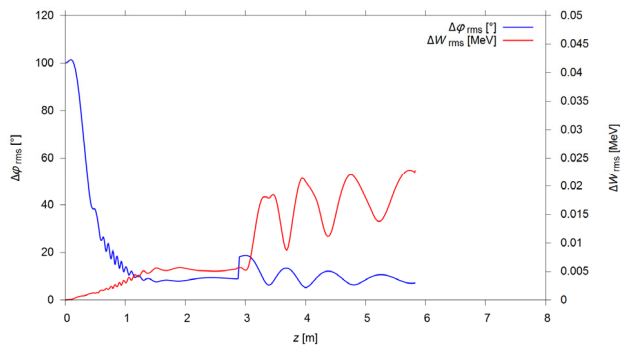

Figure 3: RMS phase and energy spread along the RFQs.

It can be seen that the beam has been gradually bunched with a relatively small energy spread until the RFJ happened at $z$ = ~3 m. Afterwards, the phase spread got an obvious jump, but it was again step by step reduced to a final size which is even smaller than that at the RFQ1 exit, with some oscillations. Though the evolution of the energy spread has a relatively larger oscillation, the absolute energy spread at the RFQ2 exit is still only ±0.6%. All these oscillations were caused by the beam mismatching at the transition.

The applied main optimization strategies for the transverse matching are:
- To use only a short drift between two RFQs.
- To remove the radial matching section of the second RFQ so that the rhythm of focusing and defocusing can be minimally disturbed.

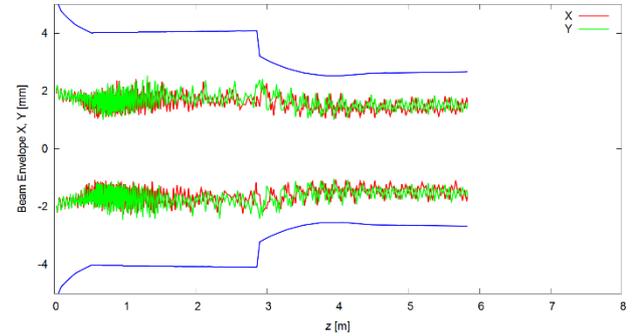

Figure 4: Transverse beam envelopes along the RFQs.

Figure 4 plots the beam envelopes in the horizontal and vertical planes along the RFQs. The beam size has been well controlled roughly within the range of ±2mm. The RFJ will shrink the electrode aperture dramatically, but the beam has still enough safety margin to the electrodes.

Figure 5 shows the influence of the RFJ more clearly. At the beginning of the RFQ2, both of the transverse emittances have a quick growth, especially in the $x$ plane where the beam from the RFQ1 is defocused (see Fig. 6). Because $\varepsilon_x$ is a little smaller than $\varepsilon_y$ at the RFQ1 output, finally we got similar transverse output emittance values at the end of the whole accelerating channel.

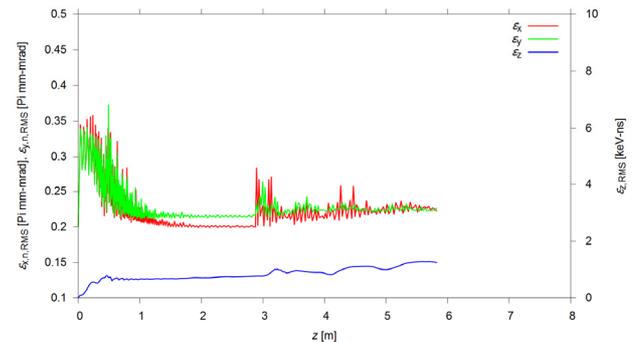

Figure 5: Emittance as a function of position.

The input and output phase spaces of the two RFQs are plotted in Fig. 6. The most remarkable result due to the RFJ

is a big phase spread jump. By means of the beam matching methods mentioned above, the main beam is confined within the area ±30° and ±0.1 MeV in the longitudinal output phase space, although there are some halo particles.

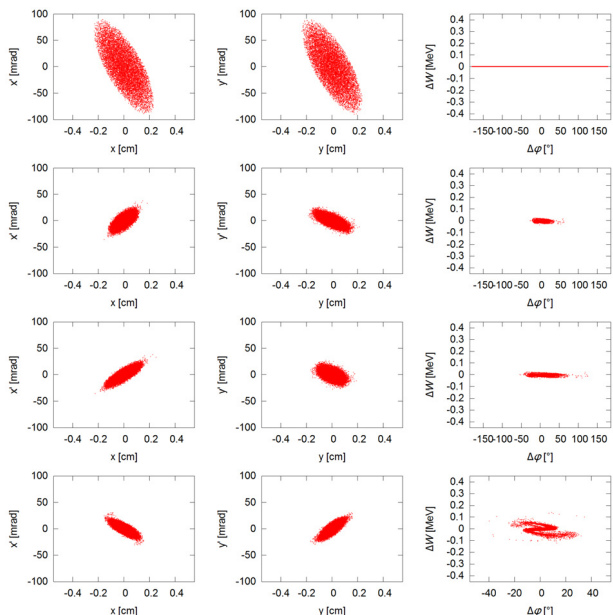

Figure 6: Particle distributions at RFQ1 input (top), RFQ1 output (2nd row, before RFJ), RFQ2 input (3rd row, after RFJ), RFQ2 output (bottom).

This RFQ-RFQ combination has been also tested with different beam currents up to 20 emA but at the same input emittance that should be still reasonable for such currents. In Fig. 7, the evolution curves of the transmission efficiency for both nominal and tested cases show satisfying results.

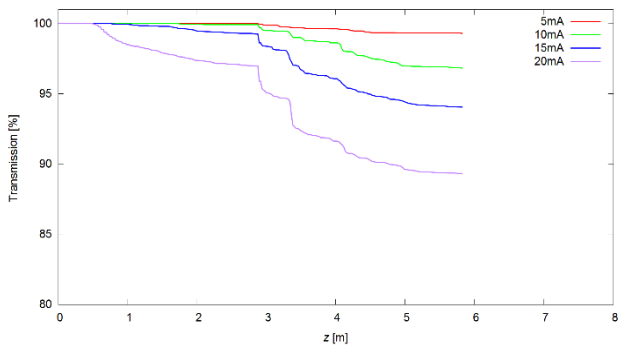

Figure 7: Beam transmission efficiency at different input currents.

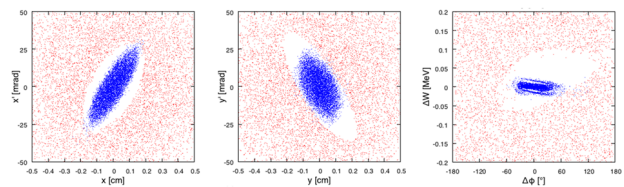

Figure 8: Acceptance plots of the RFQ2.

In Fig. 8, the empty place is the transverse and longitudinal "acceptance" of the RFQ2 and the blue dots stand for the particles injected into the RFQ2, respectively. One can see that:

- Longitudinally, the RFQ2 has still enough space to accept a larger beam. It will be even better if a beam slightly focused in the longitudinal plane will be injected.
- Transversely, the RFQ2 can already provide no further safety margin to the beam, especially in the $x$ plane where the beam is defocused.

## CONCLUSION

An RFJ between RFQs without debunching was investigated. It is feasible but challenging. A further study introducing a quadrupole lens as the MEBT to improve the transverse beam matching and the transverse "acceptance" is ongoing.

## ACKNOWLEDGEMENTS

The author C. Z. would like to acknowledge Eugene Tanke for his friendly help with the simulation code and very valuable discussions for beam matching.

## REFERENCES

[1] N. Holtkamp, "Status of the SNS Linac: An Overview", LINAC'04, Lübeck, Germany, FR103 (2004); http://www.JACoW.org

[2] Y. Liu et al., Stability Studies for J-PARC Linac Upgrade to 50 mA/400 MeV, IPAC'15, Richmond, Virginia, USA, THPF039 (2015); http://www.JACoW.org

[3] M. Eshraqi et al., "The ESS Linac", IPAC'14, Dresden, Germany, THPME043 (2014); http://www.JACoW.org

[4] Y. Yamazaki et al., "Beam Physics and Technical Challenges of the FRIB Driver Linac", IPAC'16, Busan, Korea, WEYA01 (2016); http://www.JACoW.org

[5] D. Vandeplassche et al., "The Myrrha Linear Accelerator", IPAC'11, San Sebastián, Spain, WEPS090 (2011); http://www.JACoW.org

[6] P. Krejcik, "A Cascade RFQ Principle for High Current Beams", AIP Conference Proceedings 139, 179 (1986); https://doi.org/10.1063/1.35573

[7] C. Zhang, H. Podlech, Nucl. Instrum. Methods Phys. Res., Sect., A 879 (2018) 19–24.

[8] T.P. Wangler, *RF Linear Accelerators*, (Wiley-VCH Verlag GmbH & Co. KG, 2008).

[9] C. Zhang et al., Phys. Rev. ST Accel. Beams 7, 100101 (2004).

[10] C. Zhang, A. Schempp, Nucl. Instrum. Methods Phys. Res., Sect., A, 586 (2008) 153-159.

[11] http://dynac.web.cern.ch/dynac/dynac.html

[12] E. Tanke, et al., "DYNAC: A Multi-Particle Beam Dynamics Code for Leptons and Hadrons in Complex Accelerating Elements", LINAC'02, Gyeongju, Korea, TH429 (2002); http://www.JACoW.org